\documentstyle[11pt]{article}

\newcommand{\s}{\sigma}

\def\beq{\begin{equation}}
\def\eeq{\end{equation}}

\def\Im{\rm Im}
\def\ct{\tilde c_t}

\def\gam{\gamma}

\def\lam{\lambda}

\newcommand{\CL}{{\cal L}}

\newcommand{\pl}[3]{, Phys.\ Lett.\ {{\bf #1}} {(#2)} {#3}}
\newcommand{\np}[3]{, Nucl.\ Phys.\ {{\bf #1}} {(#2)} {#3}}
\newcommand{\pr}[3]{, Phys.\ Rev.\ {{\bf #1}} {(#2)} {#3}}
\newcommand{\prl}[3]{, Phys.\ Rev.\ Lett.\ {{\bf #1}} {(#2)} {#3}}

\newcommand{\zp}[3]{, Z.\ Phys.\ {{\bf #1}} {(#2)} {#3}}

\begin{document}
\begin{titlepage}
\begin{flushright}
\hfill{YUMS 98-001}\\
\hfill{SNUTP 98-001}\\
\end{flushright}
\vspace{2.0cm}
\begin{center}
{\Large\bf $T$-ODD GLUON--TOP-QUARK COUPLINGS AT LHC\footnote{Talk by 
Jake Lee at the APCTP workshop: ``Pacific Particle Physics Phenomenology"
(Oct 31--Nov 2, 1997).}}\\
%\hfill{}
\vskip 2cm
{JAKE LEE, S.Y. CHOI and C.S. KIM}\\
\vskip 0.5cm
{\sl Department of Physics, Yonsei University, Seoul 120-749, Korea}
\end{center}

\vskip 2cm
\setcounter{footnote}{0}

\begin{abstract}
The $T$--odd top--quark chromoelectric dipole moment, $d_t$, 
is probed through top--quark--pair production via gluon fusion 
at the CERN LHC with the possibility of having polarized proton 
beams in account. At 1-$\sigma$ level, the typical
$CP$-odd lepton energy and tensor correlations enable us to
measure $Re(d_t)$ and $Im(d_t)$ up to the order of
10$^{-18}$ ($g_s{\rm cm}$) in the unpolarized case and 
the initial $CP$-odd gluon spin correlation allows us to probe 
$Im(d_t)$ up to the order of 10$^{-20}$ 
($g_s{\rm cm}$) for $\sqrt{s}=14$ TeV and the 
integrated luminosity ${\cal L}_{pp}=10$ fb$^{-1}$.
\end{abstract}

\end{titlepage}

\noindent
Top quark is expected to be sensitive to new physics at TeV energy scale,
not readily observable in lighter quarks, and it is copiously produced at
proton-proton collisions at LHC, predominantly through gluon--gluon fusion.
Therefore, we probe the possibility of measuring the $T$-odd basic coupling of
gluon and top quark, top-quark chromoelectric dipole moment ($t$CEDM) 
at LHC~\cite{jake}.

The process $gg\rightarrow t\bar{t}$, followed by $t$ and 
$\bar{t}$ semileptonic decays, has been studied to extract the real part 
of the $T$-odd $t$CEDM form factor by using the optimal observables~\cite{atwd}.
We extend the work to measuring the imaginary part of $t$CEDM 
and extract the real and imaginary parts through 
typical $CP$-odd lepton and antilepton correlations~\cite{bern} in the
$t$ and $\bar{t}$ semileptonic decays.

It has been demonstrated~\cite{guni} that the $CP$ properties of the Higgs 
boson can be directly probed with the initial $CP$-odd spin correlations 
of gluons, whose polarization is from polarized proton beams.  
We apply the initial gluon-gluon spin correlations to measuring 
the imaginary part 
of the $t$CEDM and compare them with the $CP$-odd lepton correlations 
from the $t$ and $\bar{t}$ semileptonic decays.

In general the gluon--top--quark interaction Lagrangian consist of 
two Standard Model(SM) dimension-four terms and two dimension-five terms:
\begin{eqnarray}
\CL_M={1\over 2}g_s\left({c_t\over 2m_t}\right){\bar t}
        \s^{\mu\nu}G^a_{\mu\nu}T_a t,\qquad 
\CL_E={i\over 2}g_s\left({\ct\over 2m_t}\right){\bar t}
        \s^{\mu\nu}\gamma_5 G^a_{\mu\nu}T_a t,
\end{eqnarray}
where $\s^{\mu\nu}={i\over 2}[\gam^\mu,\gam^\nu]$, $G^a_{\mu\nu}$ is the gluon
field strength, and $T_a={1\over 2}\lam_a$ ($a=1$ to $8$).
Among the four terms, only the $\CL_E$ violates $T$ invariance and 
defines the $t$CEDM $d_t\equiv g_s({\ct/2m_t})$. In the SM, 
this $t$CEDM form factor is extremely small~\cite{dono} because it arises 
only at three or more loops.  On the other hand, the $t$CEDM can be much 
larger in many models of $CP$ violation such as the multi-Higgs-doublet models
and Minimal Supersymmetric Standard Model~\cite{sony}.
In this light, a nonvanishing $t$CEDM should be a strong 
indication of new physics beyond the SM.

Firstly, let us consider the case that proton beams are unpolarized.
In this case the initial unpolarized 
gluon-gluon system in the process $gg\rightarrow t\bar{t}$ is $CP$ invariant 
and therefore any detection of the $t$CEDM requires information on the final 
$t\bar{t}$ spin correlations. The spin correlations can be indirectly inferred 
through the lepton and antilepton correlations of the semileptonic top and 
antitop quark decays $t\rightarrow bl^+\nu_l$ and 
$\bar{t}\rightarrow bl^-\bar{\nu}_l$ ($l=e,\mu$).
We may employ two typical $CP$-odd 
and $CP\tilde{T}$-even correlations~\cite{bern} 
\begin{eqnarray}
A_1={\hat p}_g\cdot({\vec p}_{\bar l}\times {\vec p}_l),\qquad
T_{33}=2({\vec p}_{\bar l}
      -{\vec p}_l)_3({\vec p}_{\bar l}\times{\vec p}_l)_3,
\end{eqnarray}
and three $CP$-odd and $CP\tilde{T}$-odd correlations 
\begin{eqnarray}
&& A_E=E_{\bar l}-E_l,\qquad 
   A_2={\hat p}_g\cdot({\vec p}_{\bar l}+ {\vec p}_l),\nonumber\\
&& Q^l_{33}=2({\vec p}_{\bar l}+{\vec p}_l)_3({\vec p}_{\bar l}-{\vec p}_l)_3
      -\frac{2}{3}({\vec p}^2_{\bar l}-{\vec p}^2_l).
\end{eqnarray}
However, Bose symmetry of the initial $gg$ system forces 
the vector correlations $A_1$ and $A_2$ to vanish so that no information
on the $t$CEDM can be extracted from them. So, we should use $T_{33}$ 
as a $CP\tilde{T}$-even correlation, which is proportional to the real part
of the $t$CEDM, and $A_E$ and $Q_{33}$ as $CP\tilde{T}$-odd correlations,
which is proportional to the imaginary part of the $t$CEDM.

In our numerical analysis we use the following experimental parameters: 
\begin{eqnarray}
&& \epsilon=10\%,\qquad \hskip 0.6cm 
   B_l=B_{\bar{l}}=21\% \ \ {\rm for}\ \ {l=e,\mu},\nonumber\\
&& \sqrt{s}=14\ \ {\rm TeV},\ \  
   \CL_{pp}=10\ \ {\rm fb}^{-1},\ \
   m_t=175\ \ {\rm GeV},
\label{exp_para}
\end{eqnarray}
where $\epsilon$ stands for a detection efficiency.
For the unpolarized gluon distribution function, we employ the GRVHO 
parametrization~\cite{grv}.
\begin{table}[t]
\centering
\caption{Attainable 1-$\sigma$ limits on $Re(d_t)$ and $Im(d_t)$, through  
         $T_{33}$, $A_E$ and $Q_{33}$ for the parameter set (\ref{exp_para}).}
\vspace{0.4cm}
\begin{tabular}{|c|c|}
%\hline\multicolumn{4}{c}{$\sigma \, (nb)$ \qquad $p_{_T} >10 \, GeV$,
%                   \quad $ |y| < 2.5 $}\\ \hline\hline
\hline
     observable & Attainable 1-$\sigma$ limits \\ \hline
       $T_{33}$ & $|Re(d_t)|=0.899\times 10^{-17} g_s{\rm cm}$\\
       $A_E$    & $|Im(d_t)|=0.858\times 10^{-18} g_s{\rm cm}$\\
       $Q_{33}$ & $|Im(d_t)|=0.205\times 10^{-17} g_s{\rm cm}$\\ 
        \hline
\end{tabular}
\end{table}
Table~1 shows the 1-$\sigma$ sensitivities of the $CP$-odd correlations, 
$T_{33}$, $A_E$ and $Q_{33}$ to $Re(d_t)$ and $Im(d_t)$, respectively, 
for the parameter set (\ref{exp_para}). 
Quantitatively, $T_{33}$ and $Q_{33}$ enable us to probe $Re(d_t)$ and
$Im(d_t)$ of the order of $10^{-17}g_s {\rm cm}$, respectively, and 
$A_E$ allows us to probe $Im(d_t)$  down to the order of 
$10^{-18}g_s {\rm cm}$. 

Secondly, let us consider the case that proton beams are polarized.
It has been claimed in many works~\cite{brod} that gluons in a polarized proton should
be polarized to explain the observed EMC effect. We assume in this
work that the polarization transmission from protons
to gluons exist and use the gluon polarization to form a
$CP$-odd gluon-gluon spin correlation to measure the $t$CEDM.
This initial $CP$-odd configuration allows us to consider all 
the detectable $t\bar{t}$ decay
modes without any full reconstruction of the decay products. 
One of the simplest $CP$-odd asymmetries is the rate asymmetry:
\begin{eqnarray}
A\equiv\frac{\s_+-\s_-}{\s_++\s_-},
\end{eqnarray}
which has been used to probe the $CP$ properties of the Higgs boson~\cite{guni}.
Here, $\s_\pm$ is the cross section of $t\bar t$ production in collision of 
an unpolarized proton and a proton of helicity $\pm$.
It is straightforward to check that the difference and sum of the two 
cross sections for opposite proton helicities are given by
\begin{eqnarray}
d\s_+-d\s_- \sim g_1\Delta g_2Im(\tilde{c}_t), \qquad 
d\s_++d\s_- \sim g_1 g_2 ,
\end{eqnarray}
where $g_1(g_2)$ is the helicity-summed gluon distribution function inside 
the unpolarized (polarized) proton, $g_{1,2}=g_{1,2+}+g_{1,2-}$, and 
$\Delta g_2=g_{2+}-g_{2-}$, where $\pm$ is for gluon helicity.

The sensitivity of the $CP$-odd rate asymmetry to $Im(d_t)$ is crucially
dependent on the degree of gluon polarization achievable for polarized protons
at the CERN LHC. At present the function $\Delta g(x)=g_+(x)-g_-(x)$ 
indicating the degree of gluon polarization in a polarized proton 
is not precisely known except for the fact that it should satisfy the 
asymptotic boundary conditions: $\Delta g(x)/g(x)\rightarrow 1$ for 
$x\rightarrow 1$ and $\Delta g(x)/g(x)\propto x$ for $x\rightarrow 0$.
Several models~\cite{brod}, which satisfy these constraints, suggest that a
significant amount of the proton spin is carried by gluons.
In our numerical analysis we employ the BQ parametrization~\cite{berg} 
as an example of $\Delta g$. 
\begin{table}[t]
\centering
\caption{The number of $t\bar{t}$ events $N$, and the attainable 1-$\sigma$ 
         limits $|\Im(d_t)|$, for various $p_{_T}$-cut values for  
         $\sqrt{s}=14$ ${\rm TeV}$, $m_t = 175$ ${\rm GeV}$, 
         $\CL=10$ ${\rm fb}^{-1}$, and $\epsilon=10\%$.}
\vspace{0.4cm}
\begin{tabular}{|c|cc|}
\hline
 $p_{_T}$-cuts(GeV) & $N(\times 10^6)$
                    & $|Im(d_t)| (\times 10^{-20}g_s{\rm cm})$\\ \hline
        0   &  2.62  &  0.766 \\
%        20  &  2.55  &  0.788 \\
        40  &  2.36  &  0.847 \\
%        60  &  2.08  &  0.951 \\
        80  &  1.74  &  1.107 \\ \hline
%        100 &  1.41  &  1.313 \\ \hline
\end{tabular}
\end{table}

Table~2 shows the number of $t\bar t$ events, $N$, and the 1-$\sigma$ limits 
of $|Im(d_t)|$ for an integrated luminosity of $10$ ${\rm fb}^{-1}$. 
Note that there is no drastic $p_{_T}$-cut dependence. This feature can be
used to make a large $p_{_T}$ cut to reduce large background effects 
without spoiling the sensitivities of the rate asymmetry to the $t$CEDM. 
Remarkably, $Im(d_t)$ are much more strongly constrained
by the rate asymmetry than the lepton and antilepton energy correlation
$A_E$ (See Table~1.). 
Numerically, it is possible to measure $Im(d_t)$ up to the order of 
10$^{-20}$ $g_s{\rm cm}$ at 1-$\sigma$ level, although precise
limits require more precise theoretical estimates and experimental 
determinations of the polarized gluon distribution. 

To summarize, we can measure (i) the real and imaginary parts of 
the $T$-odd $t$CEDM up to the order of 10$^{-18}$ $g_s{\rm cm}$ 
by the $CP$-odd lepton energy and tensor correlations of the $t$ and 
$\bar{t}$ semileptonic decays in unpolarized proton-proton collisions, 
and (ii) the imaginary part of the $t$CEDM up to the order of 
10$^{-20}$ $g_s{\rm cm}$ through the production rate asymmetry for 
positively versus negatively polarized protons, assuming that 
the polarization transmission from the proton to gluons exists.
Certainly, the large enhancement obtained by use of proton polarization 
is worthwhile to be checked at the CERN LHC.
\section*{Acknowledgments}
JL thanks APCTP for offering the opportunity to give this talk.
The authors wish to acknowledge the financial support of Korean Research
Foundation made in the program year of 1997.

\begin{thebibliography}{99}
\bibitem{jake} S.Y.~Choi, C.S.~Kim and Jake Lee, hep-ph/9706379, to appear
 in PLB.
\bibitem{atwd} D.~Atwood, A.~Aeppli and A.~Soni\prl{69}{1992}{2754}.
\bibitem{bern} W.~Bernreuther et al.\np{388}{1992}{53} and references therein.
\bibitem{guni} J.F.~Gunion et al.\prl{71}{1993}{488}.
\bibitem{dono}E.P.~Shabalin, Sov.~J.~Nucl.~Phys.~{\bf 31}, (1980) 864.
\bibitem{sony} A.~Soni and R.M.~Xu\prl{69}{1992}{33}.
\bibitem{grv} M.~Gl\"uck, E.~Reya and A.~Vogt\zp{C53}{1992}{127}.
\bibitem{brod} S.~Brodsky and I.A.~Schmidt\pl{B234}{1990}{144}.
\bibitem{berg} E.L.~Berger and J.~Qiu\pr{D40}{1989}{778}.
\end{thebibliography}
\end{document}